# ALMA North American Integration Center Front-End Test System.

Geoffrey A. Ediss*, Joshua Crabtree, Kirk Crady, Erik Gaines, Morgan McLeod, Greg Morris, Rick Williams, Antonio Perfetto, and John Webber.

National Radio Astronomy Observatory[1], 1180 Boxwood Estate Road, Charlottesville, VA 22903, USA.
*Corresponding author:- gediss@nrao.edu, Tel 434-296-0245

*Abstract*

The Atacama Large Millimeter/submillimeter (ALMA) Array Front End (FE) system is the first element in a complex chain of signal receiving, conversion, processing and recording. 70 Front Ends will be required for the project. The Front End is designed to receive signals in ten different frequency bands. In the initial phase of operations, the antennas will be fully equipped with six bands. These are Band 3 (84-116 GHz), Band 4 (125-163 GHz), Band 6 (211-275 GHz), Band 7 (275-373 GHz), Band 8 (385-500 GHz) and Band 9 (602-720 GHz). It is planned to equip the antennas with the missing bands at a later stage of ALMA operations, with a few Band 5 (163-211 GHz) and Band 10 (787-950 GHz) receivers in use before the end of the construction project.

The ALMA Front End is far superior to any existing receiver systems; spin-offs of the ALMA prototypes are leading to improved sensitivities in existing millimeter and submillimeter observatories. The Front End units are comprised of numerous elements, produced at different locations in Europe, North America and East Asia and are integrated at several Front End integration centers (FEIC) to insure timely delivery of all the units to Chile. The North American FEIC (NA FEIC) is at the National Radio Astronomy Observatory facility in Charlottesville, Virginia, USA.

This paper describes the design and performance of the test set used at the NA FEIC to check the performance of the Front Ends, following integration and prior to shipment to Chile.

---

## I. Introduction

The Atacama Large Millimeter/submillimeter Array (ALMA), an international astronomy facility, is a partnership of Europe, North America and East Asia in cooperation with the Republic of Chile. ALMA is funded in Europe by the European Organization for Astronomical Research in the Southern Hemisphere (ESO), in North America by the U.S. National Science Foundation (NSF) in cooperation with the National Research Council of Canada (NRC) and the National Science Council of Taiwan (NSC) and in East Asia by the National Institutes of Natural Sciences (NINS) of Japan in cooperation with the Academia Sinica (AS) in Taiwan. ALMA construction and operations are led on behalf of Europe by ESO, on behalf of North America by the National Radio Astronomy Observatory (NRAO), which is managed by Associated Universities, Inc. (AUI) and on behalf of East Asia by the National Astronomical Observatory of Japan (NAOJ). The Joint ALMA Observatory (JAO) provides the unified leadership and management of the construction, commissioning and operation of ALMA.

The Array will consist of fifty 12-meter diameter antennas used for interferometry, four 12-meter diameter antennas for single-dish total power and spectroscopic observations, and twelve 7-meter diameter antennas for short spacing interferometry. All the ALMA antennas will eventually be equipped with ten frequency bands covering all atmospheric windows open from the Chajnantor site (Chile) between 30 GHz and 1 THz.

The largest single element of the Front End (FE) system is the cryostat (vacuum vessel) with the cryo-cooler attached, delivered by the Rutherford Appleton laboratory (RAL), UK. The cryostats will house the receivers, which are built as insertable cartridges and can relatively easily be installed or replaced. The corresponding warm optics, signal windows and infrared (IR) filters were delivered by the Institut de Radio Astronomie Millimétrique (IRAM), France. The operating temperature of the cryostats will be as low as 4 K (equivalent to 269 C below zero). In the initial phase of operations the antennas will be equipped with six bands. These are Band 3 (84-116 GHz, Herzberg Institute of Astrophysics, Canada), Band 4 (125-163 GHz, NAOJ, Japan), Band 6 (211-275 GHz, NRAO, USA), Band 7 (275-373 GHz, IRAM, France), Band 8 (385-500 GHz, NAOJ, Japan) and Band 9 (602-720 GHz, Netherlands Institute for Space Research (SRON), Holland). It is planned to equip the antennas with the missing bands at a later stage of ALMA operations, with a few Band 5 (163-211 GHz, Chalmers, Sweden) and Band 10 (787-950 GHz, NAOJ, Japan) receivers in use before the end of the construction project. The Local Oscillators (LO's) for all bands (except Band 5) are a NRAO responsibility and are delivered on warm cartridge assemblies (WCA's) which are attached to the bottom of the cryostat.

Full details of the Front End can be found in [1], with details of the optics given in [2]. Figure 1 is a 3D drawing of the FE showing the various components, and Figure 2 is a view of the FE on the tilt table during construction. Figure 3 shows a close up of the FE assembly with side panel removed, and Figure 4 shows cartridges for Bands 3, 6, 7, and 9 ready for insertion into the cryostat. For details of the cartridges see [3, 4, 1, and 5 respectively]. Note that the FE operates from 220 VAC 50 Hz, for operation in Chile and in order to remove the possibility of multiple ground loops the test equipment all operates at this voltage from the same motor generator set.

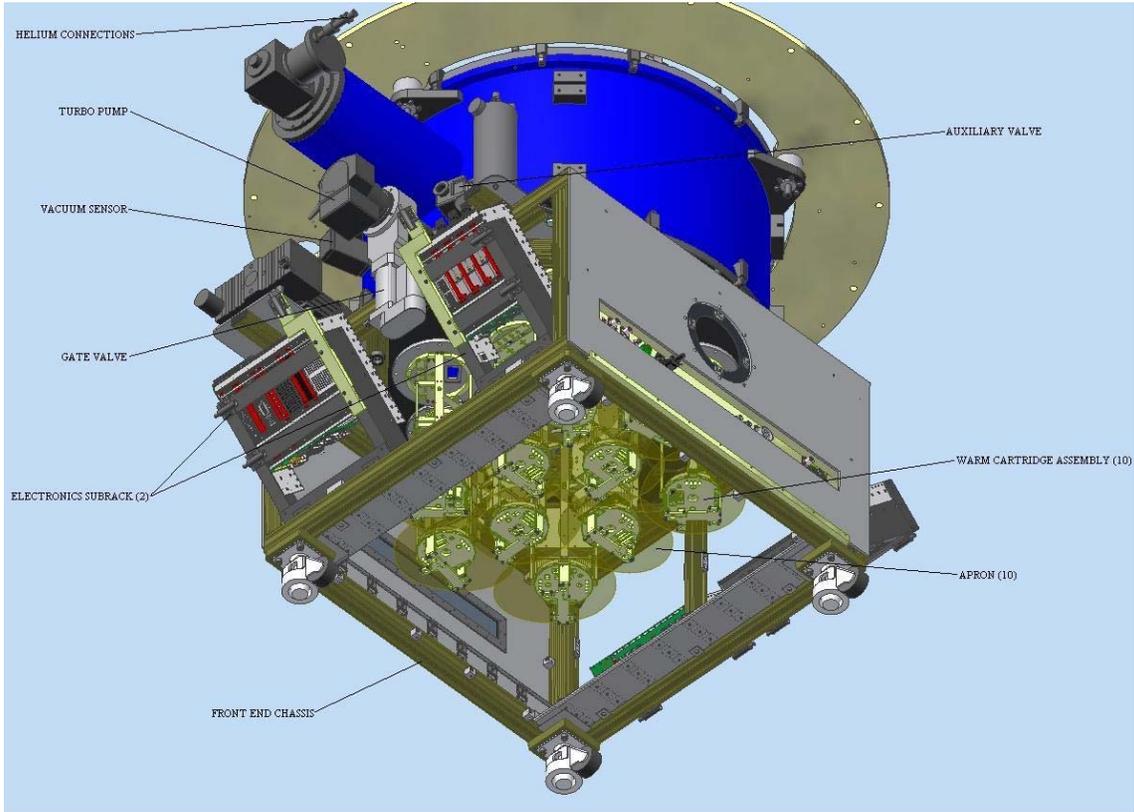

Figure 1 A 3D drawing of the Front End Assembly.

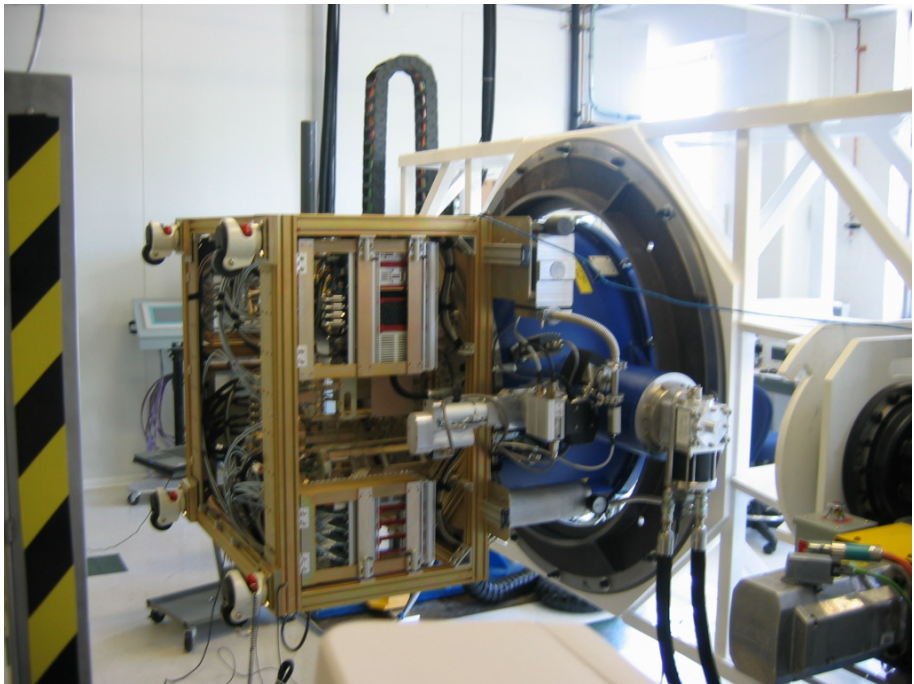

Figure 2 View of the Front End assembly on the tilt table during construction.

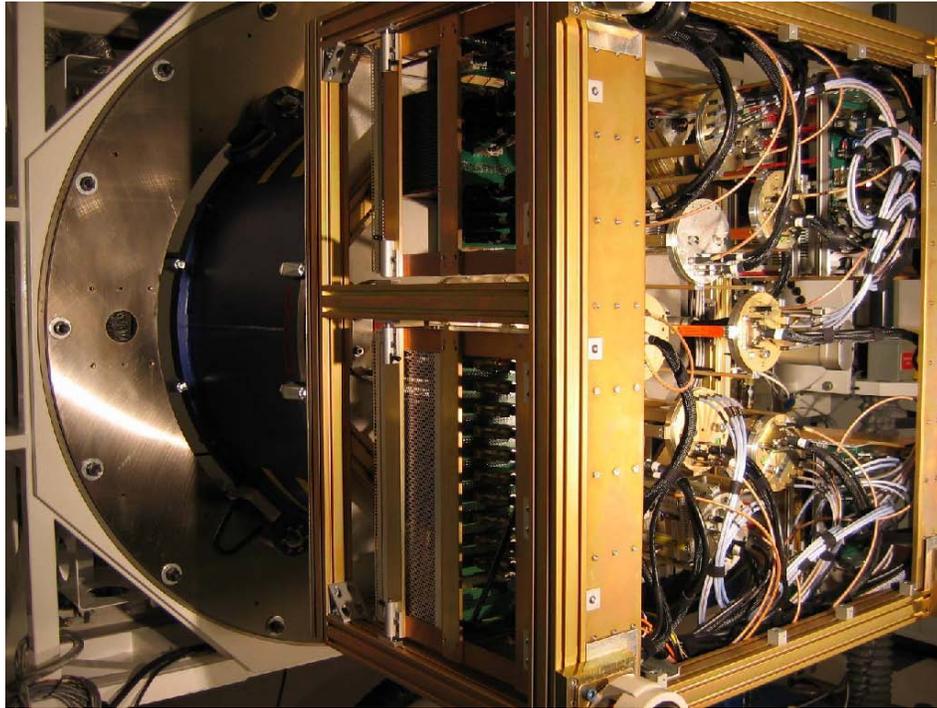

Figure 3 Close up of the FE assembly rear view with side panels removed (minus the LO Photonic Receiver).

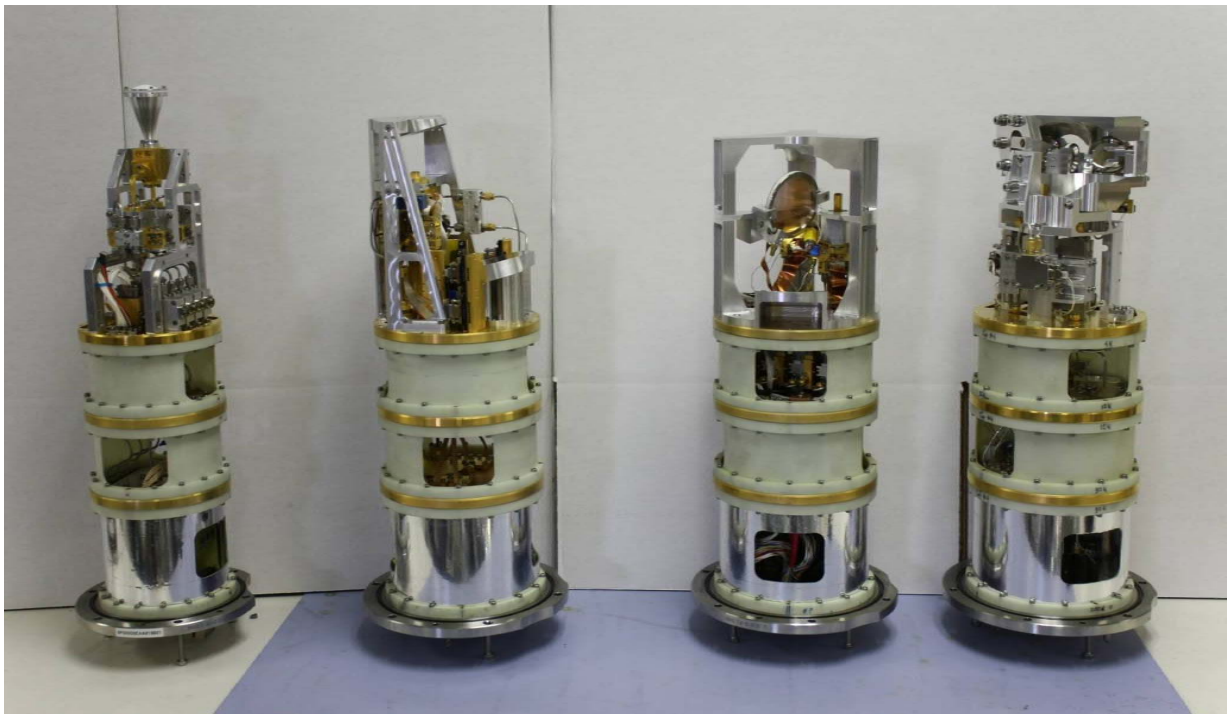

Figure 4  From the left- Bands 3 (84-116 GHz), 6 (211-275 GHz), 7 (275-373 GHz), and 9 (602-720 GHz) SIS mixer receiver cartridges, ready for insertion into the cryostat.

The ALMA Front End Integration Center test and measurement (FETMS) equipment includes the following elements:

- Environmental chamber
    This facility reproduces the thermal environment of the ALMA antenna cabin (16 to 21 C settable to +/- 0.1 C, 30-60 % relative humidity settable to +/- 2%). It houses the Front End assembly mounted on a tilt table and all ancillary equipment. It is not required to duplicate exactly the airflow through the Front End assembly as it will be at 5000m elevation, but it is designed to provide the same cooling rate using airflow at ambient pressure.

- Tilt frame
    This apparatus allows the Front End assembly to be tilted from 0-90 degrees in elevation to mimic changes in the gravity vector during telescope operation and provides 90-180 degree tilt in addition for special tests. It also carries electronics that supports the phase stability and beam measurements. The tilt frame is not required to mimic sidereal or fast-switching motion of the antenna.

- Phase measurement equipment
    The phase measurement equipment provides stable reference frequencies to which the Front End assembly and a test signal source can be phase locked. It includes equipment that can quantify the long and short term phase stability (jitter and drift) of the Front End assembly.

- Near field beam scanner
    This equipment is mounted on the tilt table above the Front End assembly and allows the determination of the beam patterns and pointing characteristics of each cartridge. The scanner source is phase locked to the phase-system reference. A laser interferometer metrology system is used to compensate for any mechanical deformation that may occur as a result of elevation changes. The scanner was manufactured by NSI [6].

- Intermediate Frequency (IF) processor
    The processor conditions the four IF outputs from the FE.

- Cold load
    The load provides a stable, low reflectivity target with an effective temperature of about 80K. A mechanized scanner mounted on the tilt frame can position the load in front of each of the cartridges.

- Monitor, control and data logging system
    Multiple applications that run the test hardware and record data into a databank are provided.

- Analysis software

Specialized software is used to analyze beam patterns, amplitude stability, gain saturation, sideband separation, etc.

As much of the measurement procedure is automated (using LabVIEW) to enable the large amounts of data to be collected in as short a time as possible (to meet the project schedule requirements). As many of the measurements as possible are automated, but it is not possible to automate changing the transmitters, which is necessary for each band.

The following table indicates the required noise temperature performance of the ALMA Front End.

**Table 1 Required Radio frequency (RF) noise temperatures.**

| Band | Single Side Band (SSB) | | Double Side Band (DSB) | |
|---|---|---|---|---|
| | $T_{SSB}$ over 80% of the RF band | $T_{SSB}$ at any RF frequency | $T_{DSB}$ over 80% of the RF band | $T_{DSB}$ at any RF frequency |
| 1 | 17 K | 26 K | NA | NA |
| 2 | 30 K | 47 K | NA | NA |
| 3 | 37 K | 60 K | NA | NA |
| 4 | 51 K | 82 K | NA | NA |
| 5 | 65 K | 105 K | NA | NA |
| 6 | 83 K | 136 K | NA | NA |
| 7 | 147 K | 219 K | NA | NA |
| 8 | 196 K | 292 K | NA | NA |
| 9 | NA | NA | 175 K | 261 K |
| 10 | NA | NA | 230 K | 344 K |

Remarks:
- In interpreting measurements, the Rayleigh-Jeans law shall be used to calculate noise temperatures.
- SSB noise temperatures shall be corrected for true single sideband response, i.e., corrected for the residual image response, which shall also be measured.
- NA means not applicable.

The Front End Integration Center test and measurement instrumentation must be capable of determining these noise temperatures to an accuracy of +- 5%. The noise performance to be measured includes contributions from warm optics, cryostat windows, and IR filters and all IF components through to the IF output ports of the Front End assembly.

Other parameters to be measured include:

Spurious response - The test and measurement instrumentation shall be able to determine the presence of spurious signals (coherent or incoherent) within the 4-12 GHz band to -40 dB per 1 MHz relative to the nominal receiver noise spectral density at the IF output.

IF output power and variations - for load temperature between 10 and 800 K at the RF input of the cartridge, the IF output power of the Front End (measured at the Front End IF outputs) must be measured to ± 1 dBm, and variations from the average IF power over the whole IF band, to an accuracy of ± 0.1 dB.

Amplitude and phase stability - the Front End amplitude stability (Allan variance, $\sigma^2(T)$) must be less than $5.0 \times 10^{-7}$ for T in the range of $0.05\ s \leq T \leq 100\ s$ and $4.0 \times 10^{-6}$ for T = 300 seconds, and rms phase noise is less than 5fs for times $20\ s \leq T < 300\ s$. The phase drift requirement refers to the 2-point standard deviation with a fixed averaging time, $\tau$, of 10 seconds and intervals, T, between 20 and 300 seconds. The system shall typically operate for at least one hour with no step discontinuities in the system delay exceeding 50 fs.

Optics measurements - The test and measurement equipment (near field beam scanner) shall be capable of measuring the Front End aperture efficiency to an accuracy of ± 2%. The scanner shall maintain this accuracy at all tilt angles. Note that the components of the aperture efficiency (taper, spillover, focus and polarization efficiencies) must be measured separately and that the polarization efficiency requires cross polarization measurements down to -30 dB. The test and measurement equipment shall be capable of measuring the direction of the E vector of the polarization channel designated "Polarization 0" to within ±0.2 degrees and the relative orientation of the E vector of the polarization channel "Polarization 0" and the E vector of the polarization channel "Polarization 1" to an accuracy of ±0.2 degrees. The system shall also be capable of measuring the co-alignment, on sky, between the beams of the orthogonal polarization channels of one cartridge to an accuracy of ±2% of the full width at half maximum (FWHM) of the primary beam.

**II. Test set design**

Overall

A top level schematic diagram of the system is given in Figure 5. It consists of the FE mounted on the tilt mechanism and the components necessary to measure:

1. Noise temperature
2. Sideband ratio
3. Beam patterns (near-field)
4. Phase stability
5. IF pass-band
6. Amplitude stability

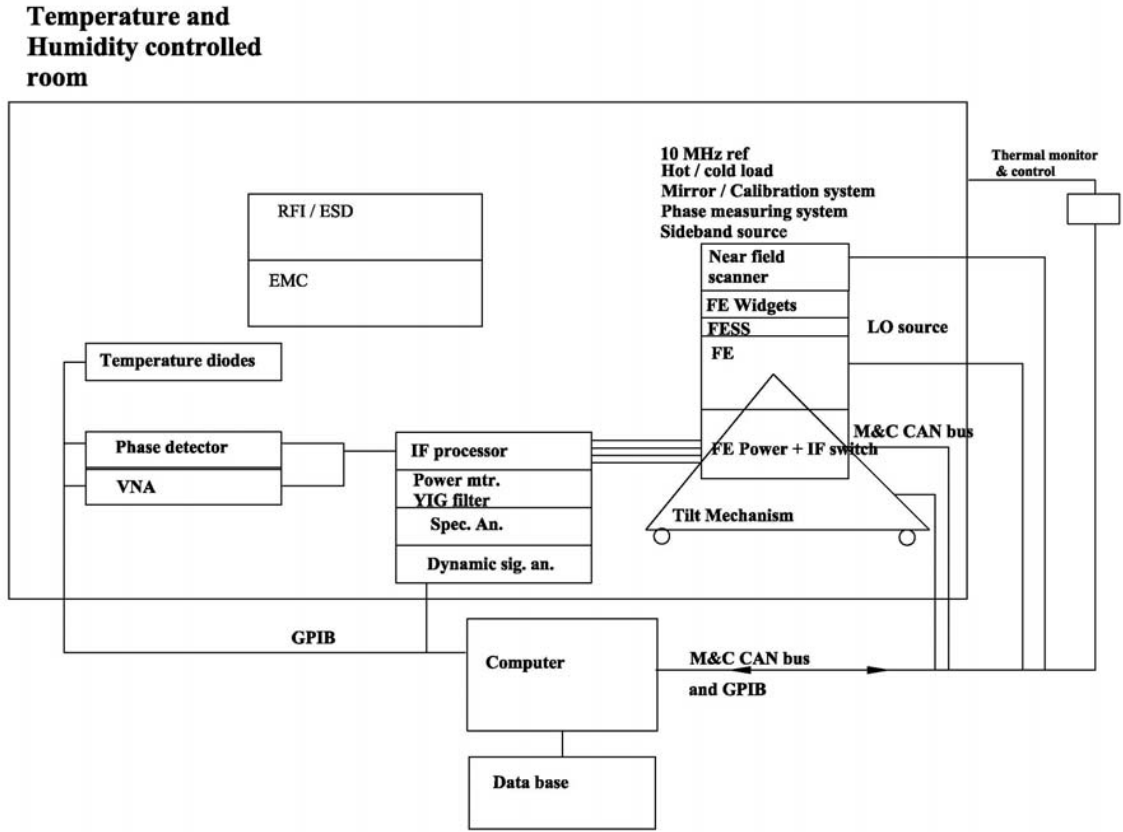

**Figure 5** A Schematic diagram of the FEIC test and measurement system.

IF processor

The IF processor amplifies, switches, filters, attenuates and measures the power of the four IF outputs from the front-end assembly. The processor must reproduce the mechanical interface to the back end.

IF processor design

The test system consists of the IF Processor chassis and its associated power supply, UPS, computer and monitor, and all additional test equipment. The additional test equipment includes: an Agilent dynamic signal analyzer (3570A), spectrum analyzer (E4408B) and quad power meters (E4418B/E4412 sensor), signal generator (E8257D), Vector network analyzer (VNA) (E8362B), Aeroflex PN 9000 Phase test set, as well as a bank of six controllers (11713A) for setting switch and attenuator configurations in the IF processor chassis.

The IF processor chassis handles four simultaneous signal paths: the upper and lower sidebands for each of two polarizations from the receiver band being measured. It was determined that a four channel processor would provide optimal data throughput versus system cost, as this configuration takes advantage of the commonality of measuring and control systems as well as several redundant signal paths. The IF processor block diagram is shown in the Figure 6, and a photograph is given in Figure 7.

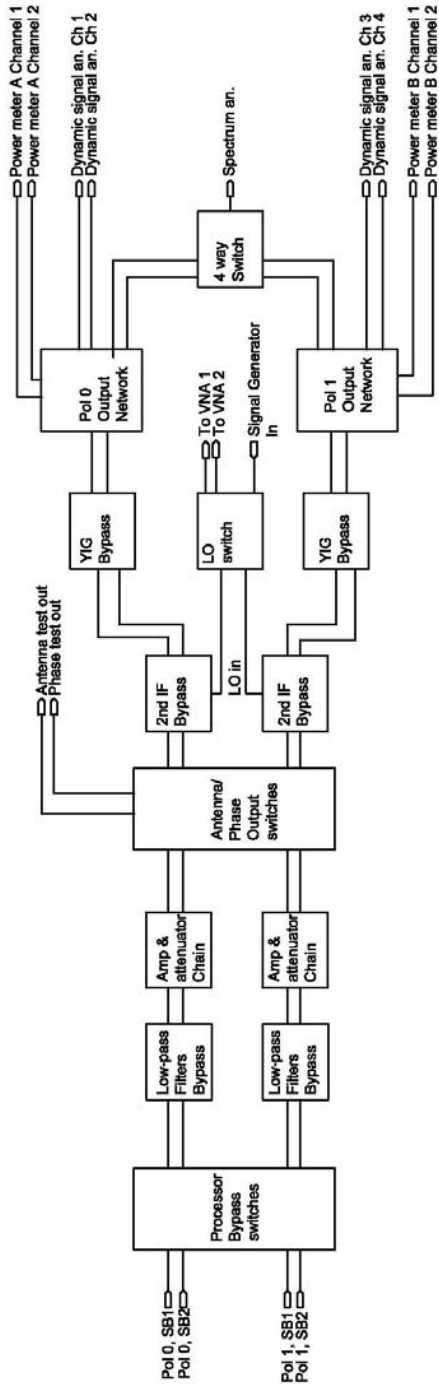

Figure 6 IF processor block diagram

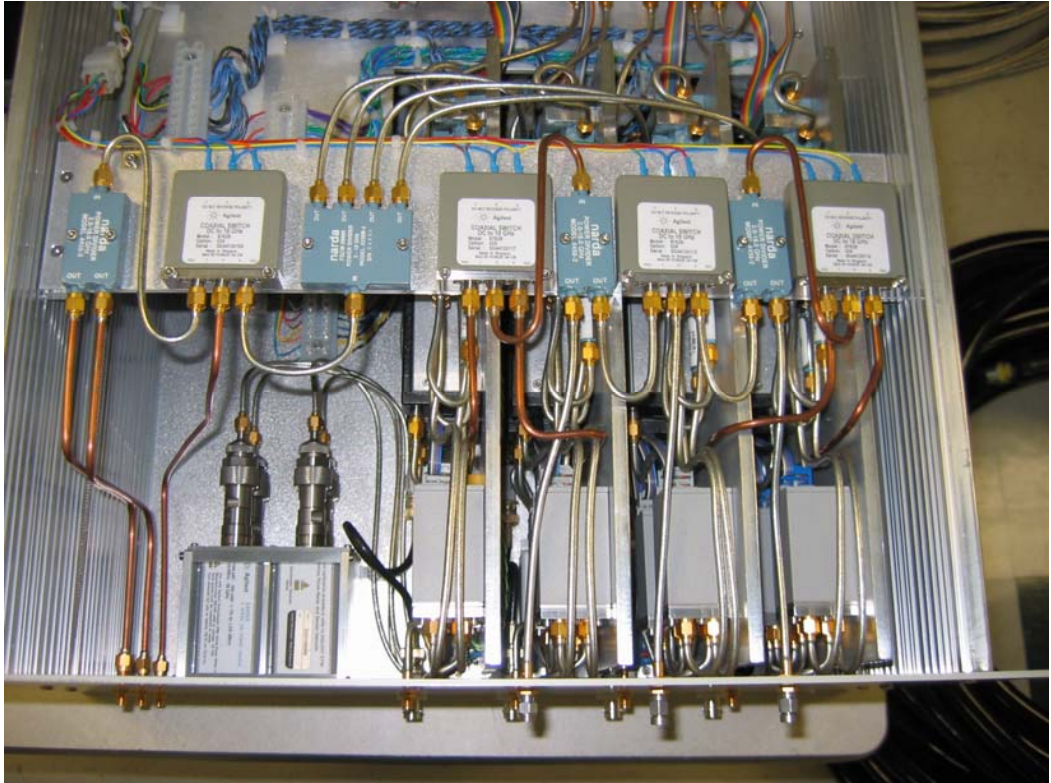

Figure 7 Internal view of the IF processor.

The signal level of each channel may be adjusted by means of 0-90 and 0-11 dB programmable attenuators to present optimum levels to the power meters (typically -30 dBm) as well as to the spectrum analyzer and the detector/dynamic signal analyzer combination.

Following IF cable and line equalizer losses, the IF processor chain typically sees an input level of approximately -59 dB. Inside the IF processor chassis, coaxial transfer switches on each plate can optionally route the input signals directly to the power meters, bypassing the processor chain. For 4-8 GHz or 4-12 GHz total power measurements, a network consisting of a high pass filter at 4 GHz and low-pass filters at 8 and 12 GHz may be switched into the signal path. An amplifier chain consisting of Quinstar amplifiers and Agilent 84904/06K attenuators provides the means of adjusting signal levels to compensate for losses associated with switching functions such as Yttrium-iron-garnet (YIG) filters or a second IF down-converter. An antenna/phase switching network allows the signal to be routed to a separate receiver system for measuring beam patterns (near-field), as well as phase stability. A 2-18 GHz mixer allows the option of converting the IF signal to a 2-GHz-wide 2nd IF for further power measurements. The local oscillator for these mixers is supplied via an Agilent E8257D signal generator. A pair of dual-tracking 4-12 GHz YIG filters may be switched into the signal path to provide a 30 MHz (3 dB bandwidth) wide signal useful for noise temperature and sideband ratio measurements. The YIG filters are digitally set via a 12-bit TTL bus, supplied by an ICS Electronics GPIB controlled parallel digital bus interface. A final switching network routes the processed signal to any of: the power meters or dynamic signal analyzer (for amplitude stability); the spectrum analyzer (for pass-band ripple, interference measurements, and general troubleshooting); or power meters (for noise temperature and sideband ratio measurements). 8743B crystal detectors inside the processor chassis and externally mounted Stanford research SR560 Low-Noise Voltage Preamplifiers provide the appropriate DC signal levels to the dynamic signal analyzer.

Figure 7 shows the internal layout of the processor, the four plates to the lower right house the four independent IF channels.

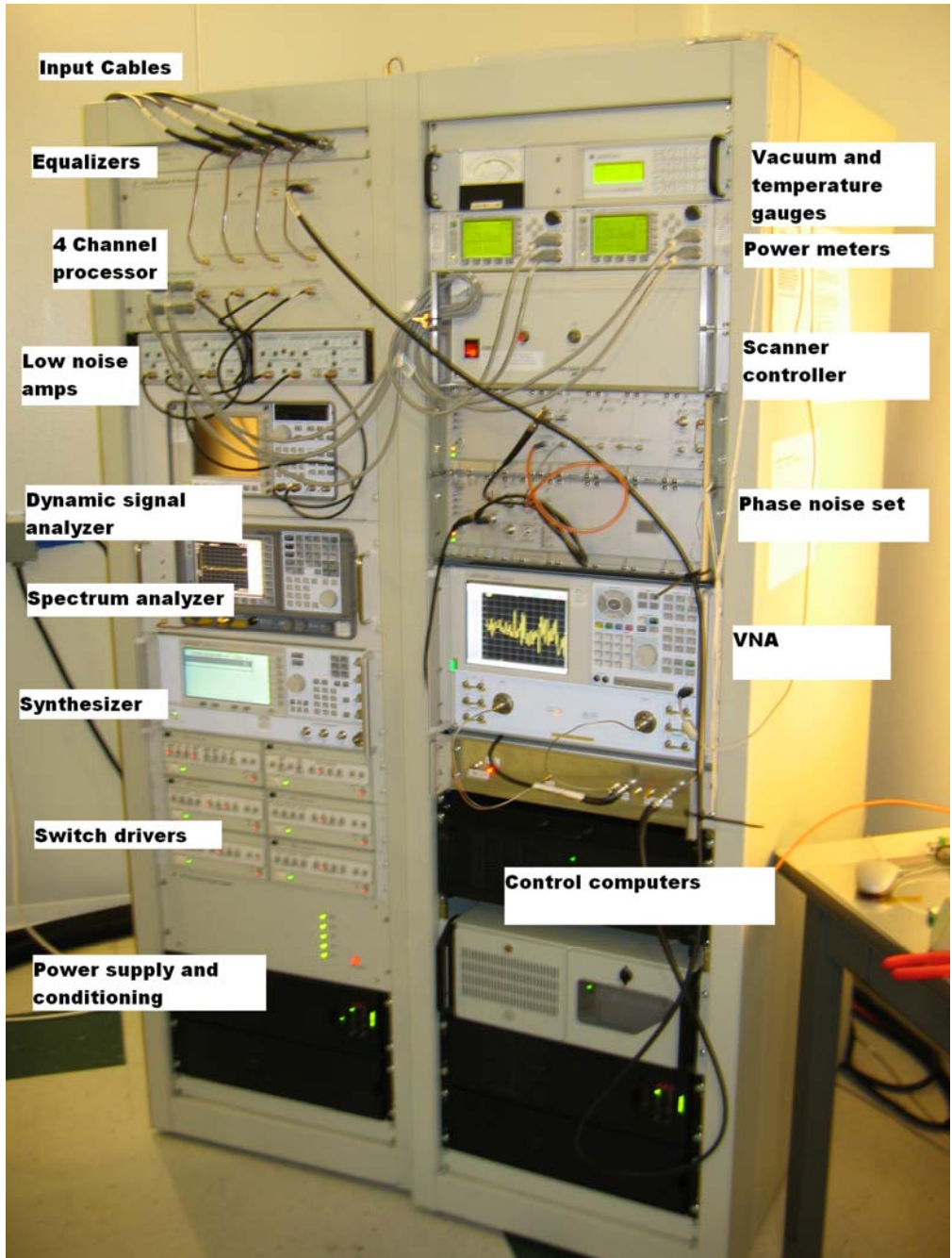

Figure 8 The IF processor racks.

The IF processor system is housed in two 19" wide x 70" high x 36" deep enclosed cabinets as shown in Figure 8. The cabinets are fan-cooled and provide an EMI-shielded cage to reduce interference from external sources, and any possible interference with the Front End. Both racks operate at 230V 50Hz and are power conditioned by an APC Smart-UPS RT5000VA RM230V. The uninterruptable power supply (UPS) provides a runtime of approximately 40 minutes during power outages, given the typical rack power consumption of 2.3 kW. This includes the power requirements of a rack-mounted computer and monitor.

Phase drift and beam scanning system

The heart of the system is the Local Oscillator Reference Test Module (LORTM). This provides phase-coherent infrared reference signals. One set is used to phase-lock the first LO of the receivers installed in the Front-end assembly, and a second set is used to phase-lock signal sources that are used to inject signals into the RF input ports of the Front End receivers. In both cases the optical reference from the LORTM is converted into a millimeter-wave signal by photo-mixers. This is required to supply the necessary signals to drive the WCA's and the beam sources in the same way as they will be driven in the telescope with the required frequency offsets.

The LORTM requires three reference frequencies for operation and these are provided by commercial synthesizers. Reference one sets the frequency difference between the reference and slave lasers. Reference two modulates the phase of the slave laser to provide sideband signals that are used by the sources. Reference three is a fixed 125 MHz signal. A 10 MHz reference is used to lock all the synthesizers together. To cover all the ALMA frequency bands, the output frequency of reference synthesizer two is modified by an array of switches, splitters, dividers and other components shown in the block diagram, Figure 9.

A VNA is used to compare the phase difference between the output of the Front End assembly and the reference signal that comes from the output of reference synthesizer #2 and display the long-term phase stability. The mixers associated with the VNA are used to eliminate drift caused by the VNA itself. The phase cancellation arrangement is illustrated in Figure 9. All the cable lengths are matched as shown to cancel thermal drifts.

An integral part of the phase measurement system are the sources that cover each of the ALMA bands (shown in Figure 10). The design of these units is based on the first local oscillators located in the warm cartridge assemblies. The higher frequency sources drive room temperature frequency multipliers followed by open-ended waveguide probes.

Database software

The database software is designed to store all FEIC test and measurement data including acceptance data of incoming subassemblies. The package also includes software that will be used to track the inbound and outbound shipping.

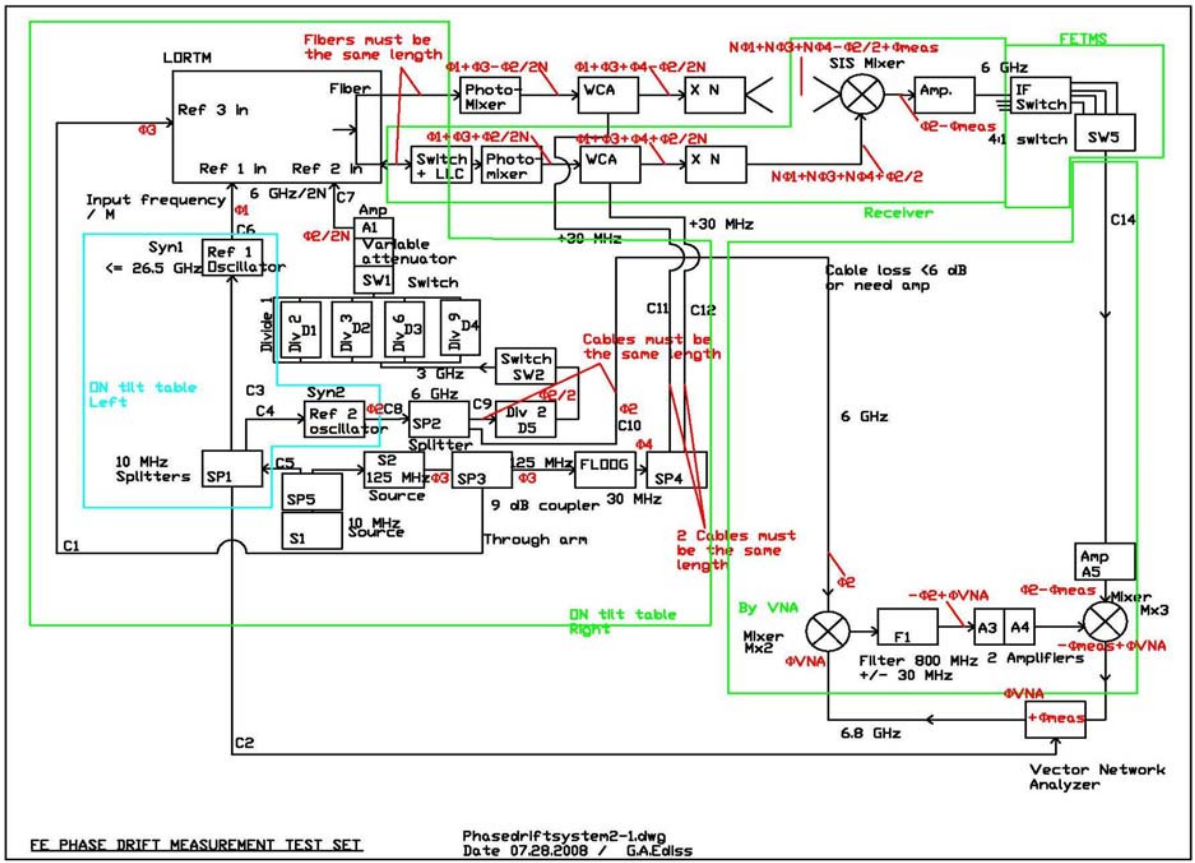

Figure 9 Block diagram of phase drift measurement system.

The FEIC database stores configuration and test data for Front Ends and Front End components produced at the FEIC. A Front End record contains information such as serial number, location, documentation links, and notes. A Component record contains serial number, location, documentation links, notes, and links to other components or Front End with which it is associated (if any). Configuration data (such as LO Parameters for a WCA) and test data (such as Noise Temperature measurements or beam patterns) will include a link to the corresponding component.

A Web-based database browser application in development allows for viewing, adding, searching, and editing records in the database. When viewing a component record, one may also view plots of associated test data, or download test data for further analysis. This tool also allows for uploading of csv files containing configuration and test data for components. Cartridge groups may upload such csv files, and the database will be automatically populated accordingly.

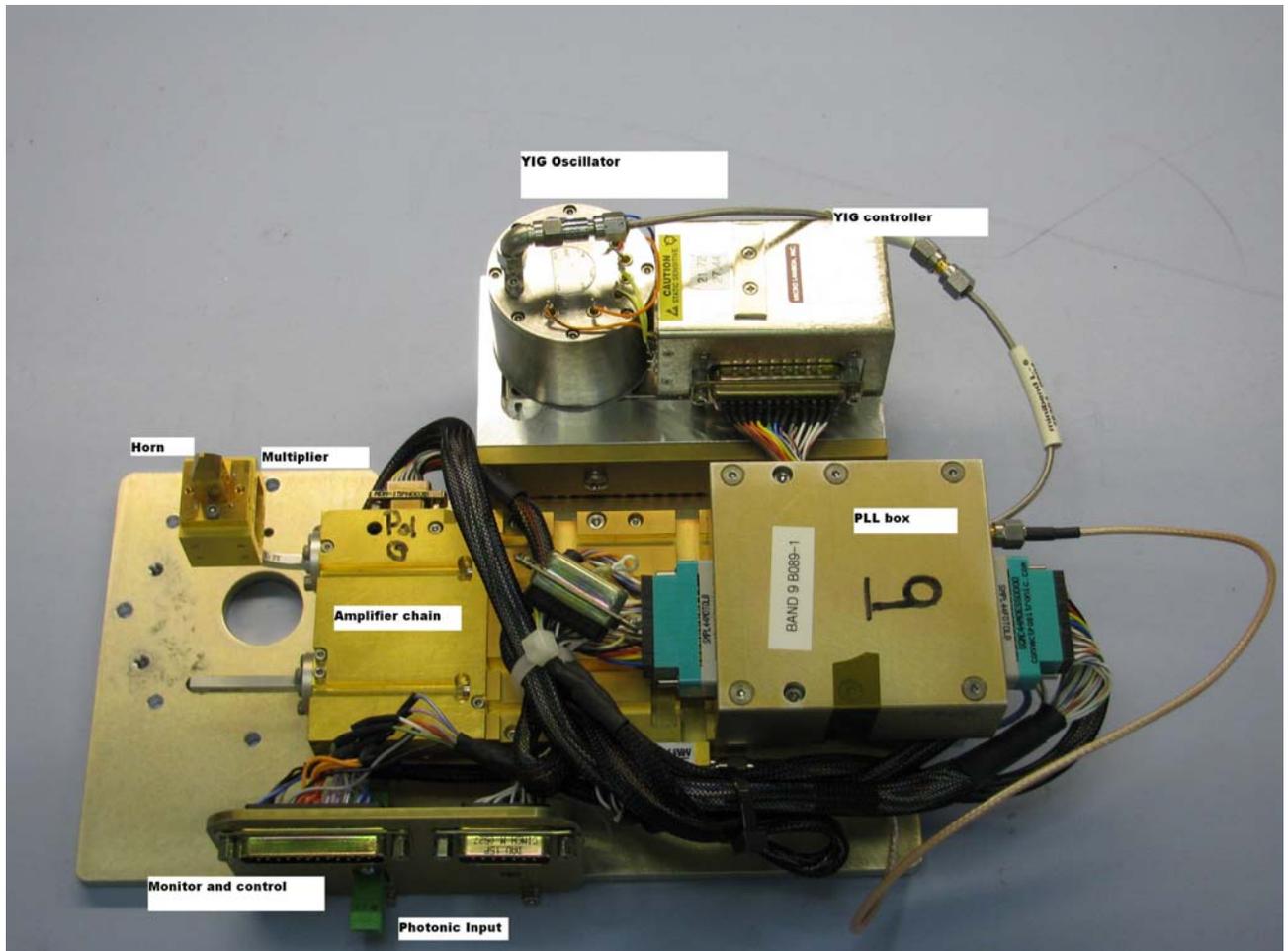

Figure 10 Modified WCA as transmitter source (Band 9). The absorber which covers the source has been removed.

Analysis software

Both the analysis and monitor & control software were written in LabVIEW 8.6, using the Endevo GOOP inheritance Toolkit. Portions of the software which control the Front End, LORTM, and other CAN devices are a C++ library. All software is designed to run on Microsoft Windows XP.

- The noise temperature calculations use the standard "Y factor" formula, corrected for sideband rejection.
- Sideband rejection is calculated according to the formula detailed in ALMA Memo 357 [7].
- The beam efficiencies are calculated using standard NSI software (for Fourier transform FFT routines, *etc*. and custom software to calculate efficiencies
- Amplitude and phase stability are calculated using Allan variance $\sigma^2(T)$

Monitor & control software

The monitor & control hardware and software applications coordinate the actions of the test and measurement system and the Front End under test, The software displays the status of the testing in progress and logs data prior to its being passed to the analysis or database software. Commercial equipment such as the tilt table, laser interferometer, LORTM and near-field beam scanner use monitor and control firmware and software written by the commercial company. In this case FETMS software is confined to interfacing, monitoring and the logging of data.

The monitor and control hardware located on the tilt table is identical to that found in a Front End assembly.

A block diagram of the FETMS hardware and associated monitor and control computers is shown in Figure 11.

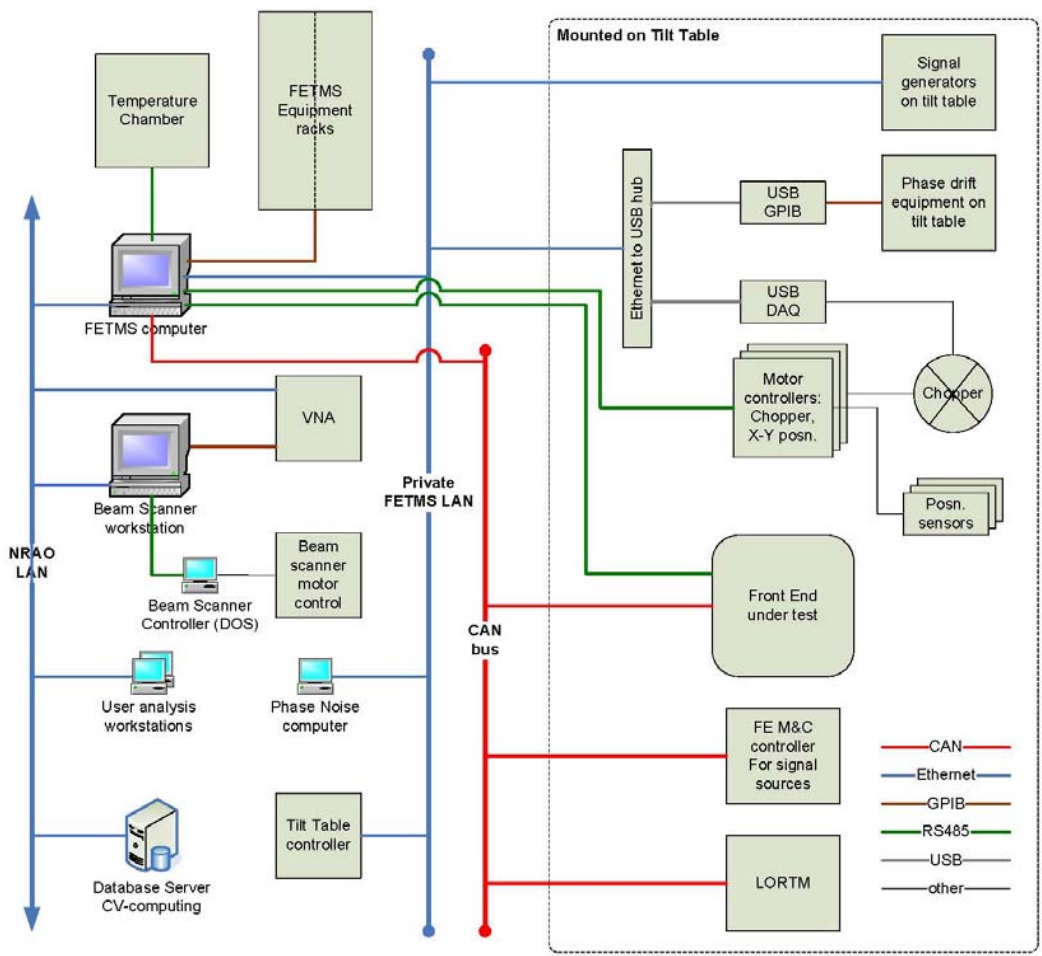

Figure 11 Block diagram of the TMS hardware and associated monitor and control computers.

Examples of the control screens are given in Figure 12 and Figure 13

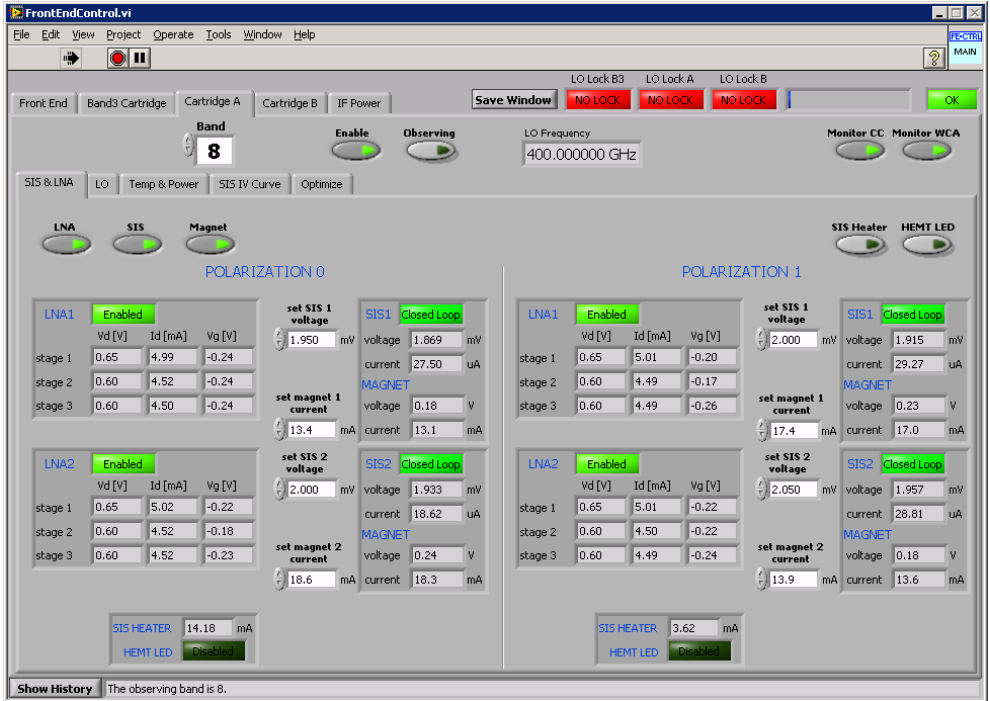
Figure 12 Front End control software screen –Band 8 set up.

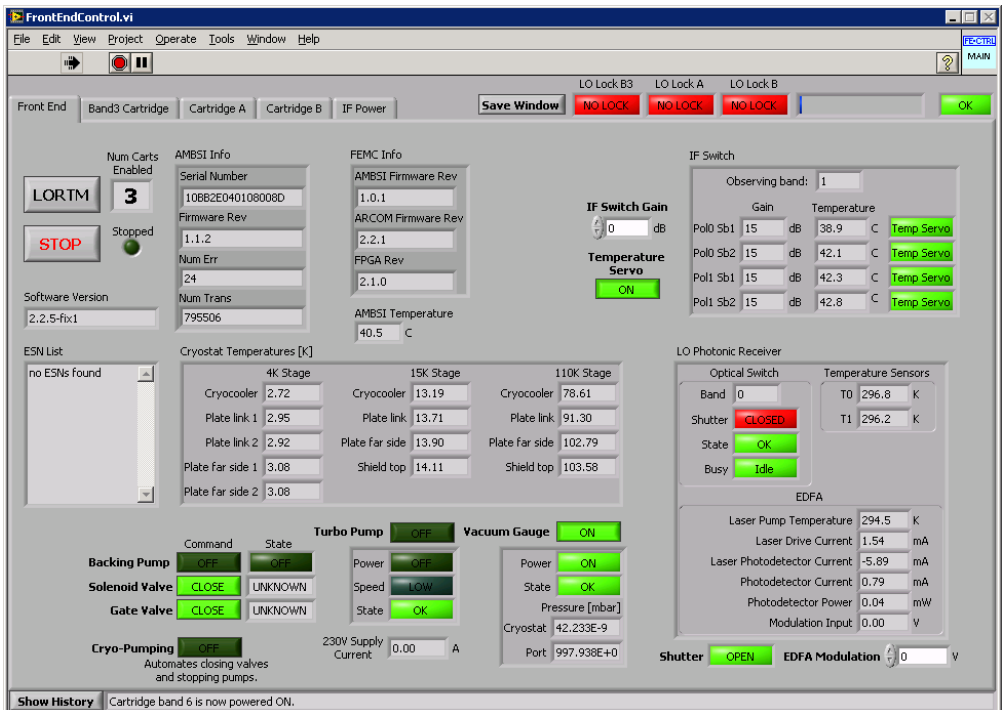
Figure 13 Front End control software screen – Cryostat temperature monitoring

Page | 16

## III. Measurements of the test set alone

The linearity and noise performance of the IF processor was checked by applying a calibrated noise source to each of the four inputs of the FETMS and varying the integral processor attenuators while sweeping the YIG filter across the 4-12 GHz band. This process was automated using a custom software application and sample results are given. Figure 14 shows the noise figure of one of the four processor channels as a function of attenuator setting and at spot IF frequencies as defined by the YIG filter. As this system follows the 80 dB of gain in the FE it will make a negligible contribution to the measured Front End noise temperature, even at the highest attenuator settings.

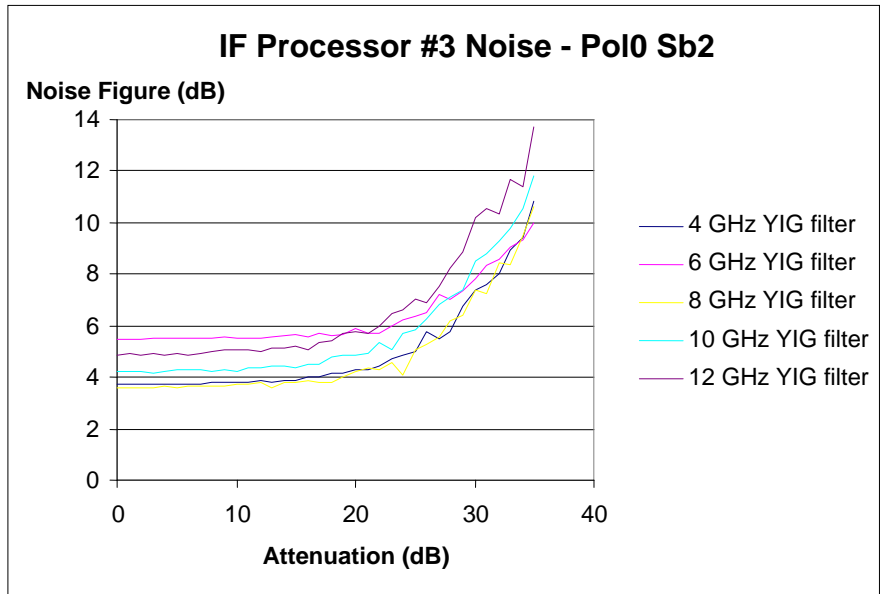

Figure 14 Channel #3 input noise.

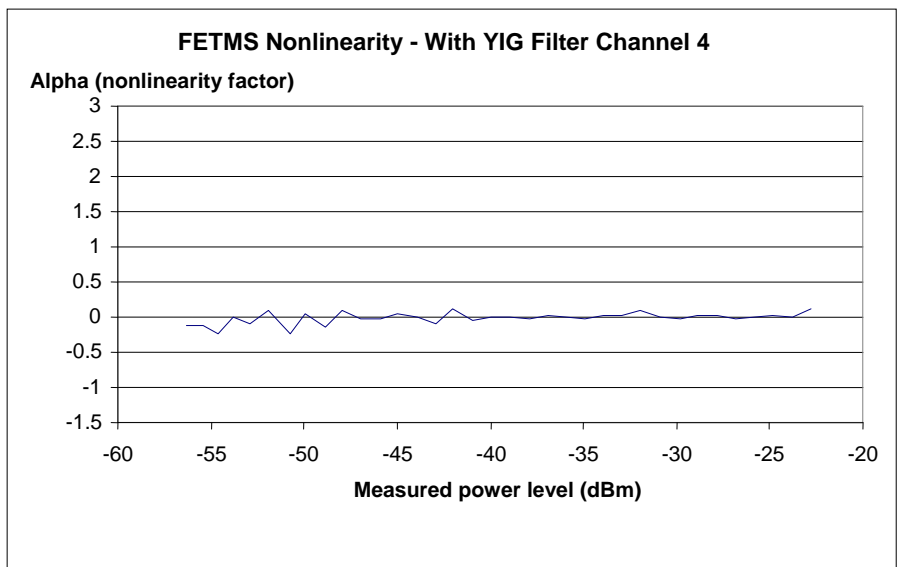

Figure 15 Non-linearity of the whole test set.

Figure 15 shows the non-linearity (expected change versus measured change for 1 dB attenuator steps, taking into account the actual attenuator step size measured with a CW source) of the test set over the normal range of use.

The spectral flatness of the IF processor and the cables connecting it to the Front End assembly was checked by installing a 50 ohm load at the end of the cables and measuring the output of the IF processor using a spectrum analyzer. The attenuators were set at their nominal values and the results over the band 4 to 12 GHz are shown in Figure 16. This shows that the equalization of the cable losses is good and that the amplifier gain varies by +/- 2 dB over the whole IF band.

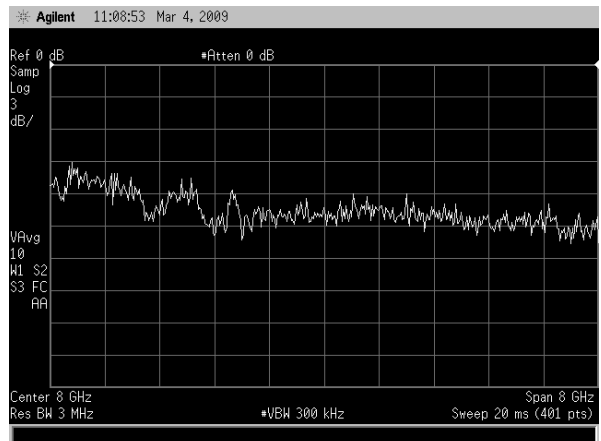

Figure 16 Pol 1 Upper sideband power over 4-12 GHz. 3 dB per vertical division.

The four channels of the IF processor (with cables and equalizers) were checked for amplitude stability by applying a 50 ohm load to each of the inputs and recording the power output using a data logger. An example of the results, shown as Allan variance is given in Figure 18 and show that the receiver specifications can be easily measured.

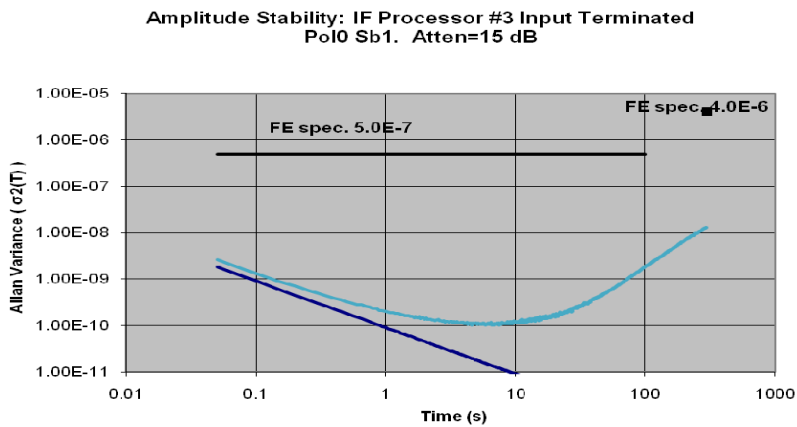

Figure 17 Amplitude stability of the test set. The horizontal black line, and point at 300 seconds is the Front End specification, and the blue line shows the slope of the radiometer equation.

Page | 18

The phase stability of the test and measurement equipment at 100 GHz was measured by comparing the output from two sources that were locked to the LORTM as shown in Figure 18, and the results using the FETMS and the phase drift cancellation scheme are given in Figure 19.

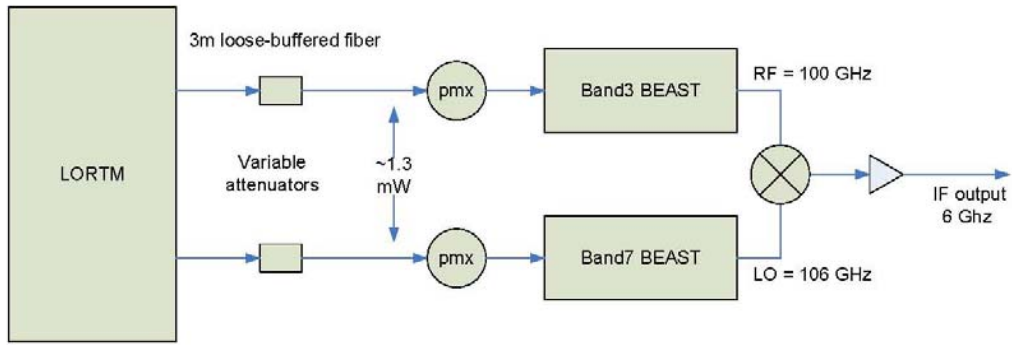

Figure 18 Phase test system using 2 Beam Scanner Test Source (Beast) units. (PMX-Phase maintaining fiber).

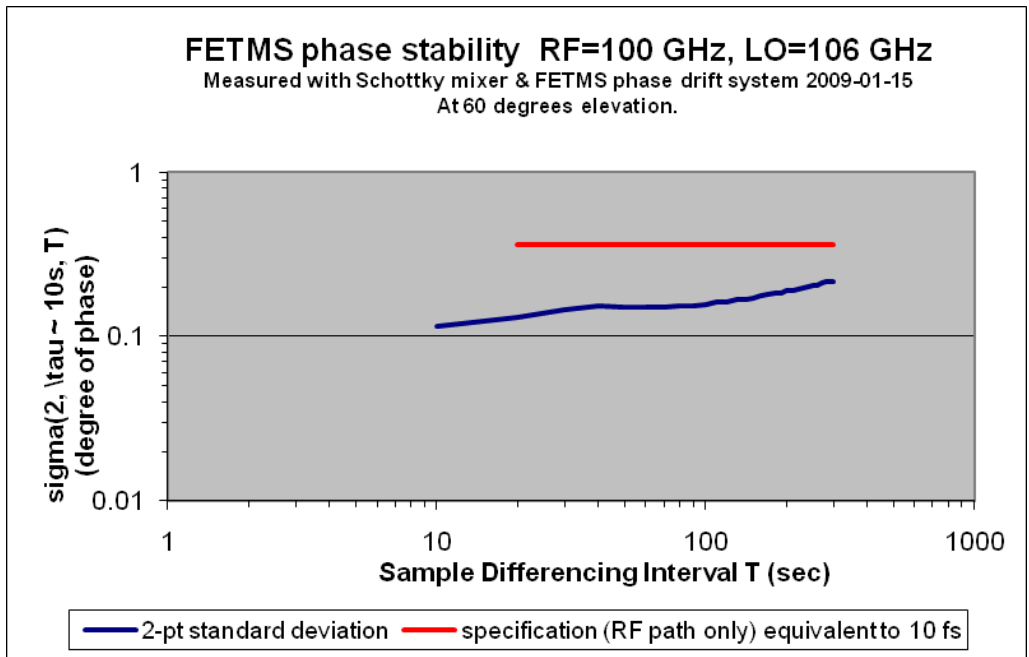

Figure 19 Phase stability of the test set.

To check the fidelity of the near-field scanner we compare the 100 GHz results from the FEIC beam scanner measurements of a test horn/lens combination (magenta) with those from the NRAO Green Bank, West Virginia far field scanner measurements of the same test horn/lens (blue). See Figure 20 for an example of the H plane cuts at 100 GHz.

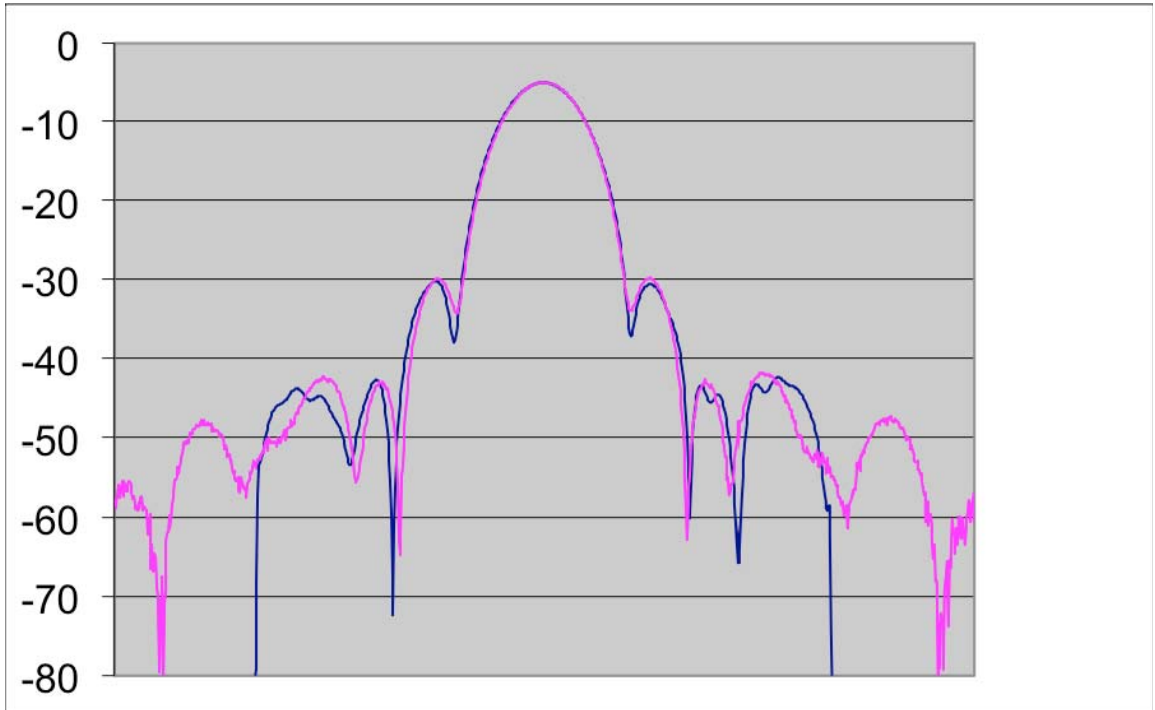

Figure 20 H plane cuts in the far field: magenta = FEIC near-field scanner, blue = Green Bank far-field scanner (due to characteristics of the far field scanner, this is not normalized to zero).

To investigate the flexure of the scanner structure as the elevation is varied we repeated the 100 GHz co-polarization beam scans at elevations of 0, 30, 45, 60, and 90 degrees. The following Table 2 details the x, y, and z position of the horn/lens waist relative to the probe (x=0, y=0, and height of the probe above the receiver horn), and the angles are the far field pointing angles in horizontal and vertical planes.

Table 2 waist position.

| Elevation | Pointing Angles (deg) | | Horn/lens to probe distance (mm) | | |
|---|---|---|---|---|---|
| | Az | El | X | Y | Z |
| 0 | 0.220 | 1.520 | -0.912 | -3.147 | -61.444 |
| 30 | 0.222 | 1.518 | -1.206 | -2.945 | -63.690 |
| 45 | 0.222 | 1.518 | -1.249 | -2.641 | -63.694 |
| 60 | 0.222 | 1.515 | -1.253 | -2.176 | -63.733 |
| 90 | 0.220 | 1.511 | -1.020 | -1.059 | -63.912 |
| 0 | 0.214 | 1.510 | -0.853 | -3.039 | -64.175 |

As this flexure is repeatable, it can be removed from the beam pointing and not affect the receiver performance on the telescope.

The isolation between the IF processor channels was measured as shown in the block diagram, Figure 22:

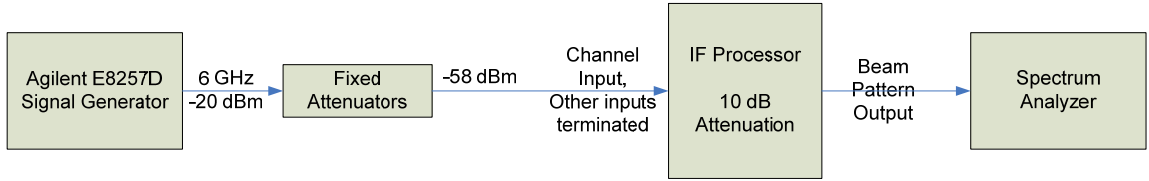

Figure 21 Block diagram of the IF processor channel isolation measurement.

The results are presented in Table 3, with a signal level of approximately -10 dBm on each channel (one at a time) the power level seen in the other channels was recorded.

Table 3 IF channel isolation

| Signal In Channel | Channel 1 (dBm) | Channel 2 (dBm) | Channel 3 (dBm) | Channel 4 (dBm |
|---|---|---|---|---|
| 1 | -10.7 | -93.6* | -89.8 | -93.8* |
| 2 | -91.8* | -12.6 | -91.6* | -92.2* |
| 3 | -95.2 | -96.3 | -10.7 | -94.2 |
| 4 | -98.0* | -98.0* | -91.9* | -11.4 |

* Noise floor

**IV. Front End results**

As proof of the operating status of the test set, a subset of measurements made from various bands of one the first Front Ends to be delivered to Chile are given in the following section.

Figure 22 shows the measured single sideband noise temperature of the band 7 receiver across the LO frequency range. The noise temperature in integrated over the 4-12 GHz IF bandwidth and is corrected for the noise injected in the image band. As can be seen, the band 7 receiver is well within the specifications.

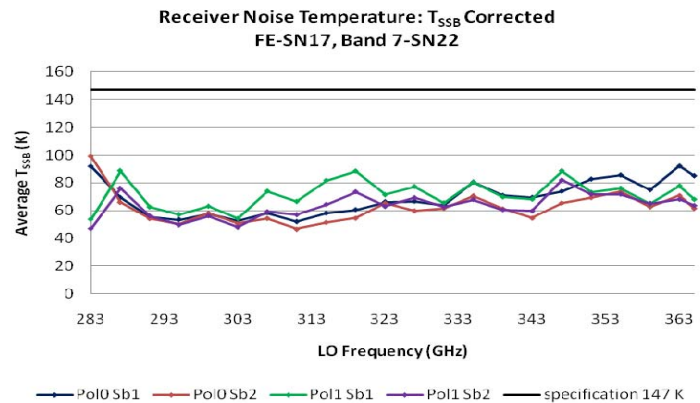

Figure 22 Band 7 receiver noise temperatures SSB (corrected for image band ratio). The horizontal line is the specification.

Page | 21

Figure 23 shows similar results for the highest band available at the moment, band 9. This band is a double sideband receiver and the noise temperatures are not corrected for the noise in the image band. Again these noise temperatures are integrated over the 4 -12 GHz IF.

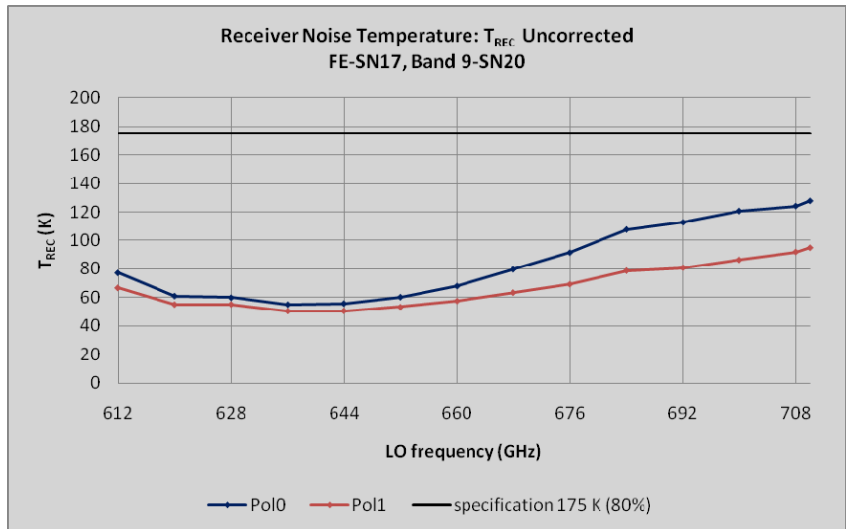

Figure 23 Band 9 receiver noise temperature (Band 9 is a double side band system) uncorrected for image band ratio. The horizontal line is the specification.

Figure 24 shows spectrum analyzer sweeps over the 4 -12 GHz IF bands for LO's between 612 and 710 GHz. This shows that no spurious signals from the LO system are seen within the resolution bandwidth of the spectrum analyzer (3 MHz).

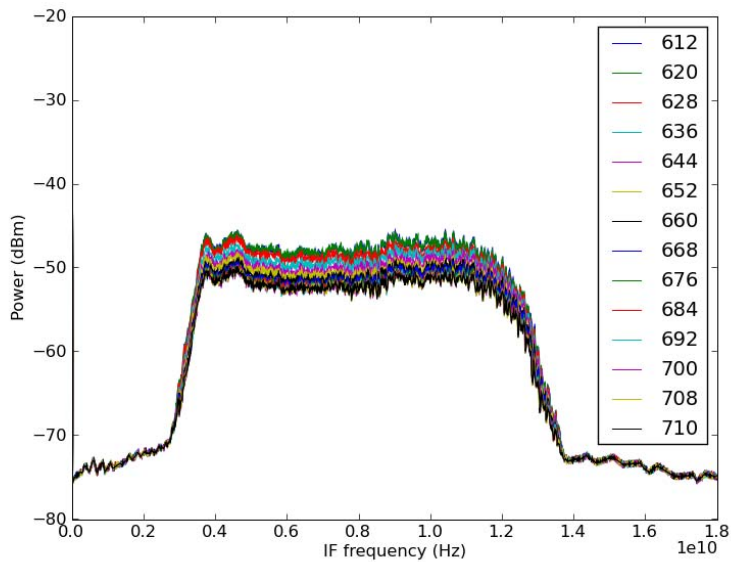

Figure 24 A sweep across the Band 9 IF showing there are no spurious signals seen.

Figure 25 gives the measured amplitude stability (Allan variance) of one polarization channel of the band 9 receiver at 662 GHz, with a 300 K absorber in-front of the receiver.

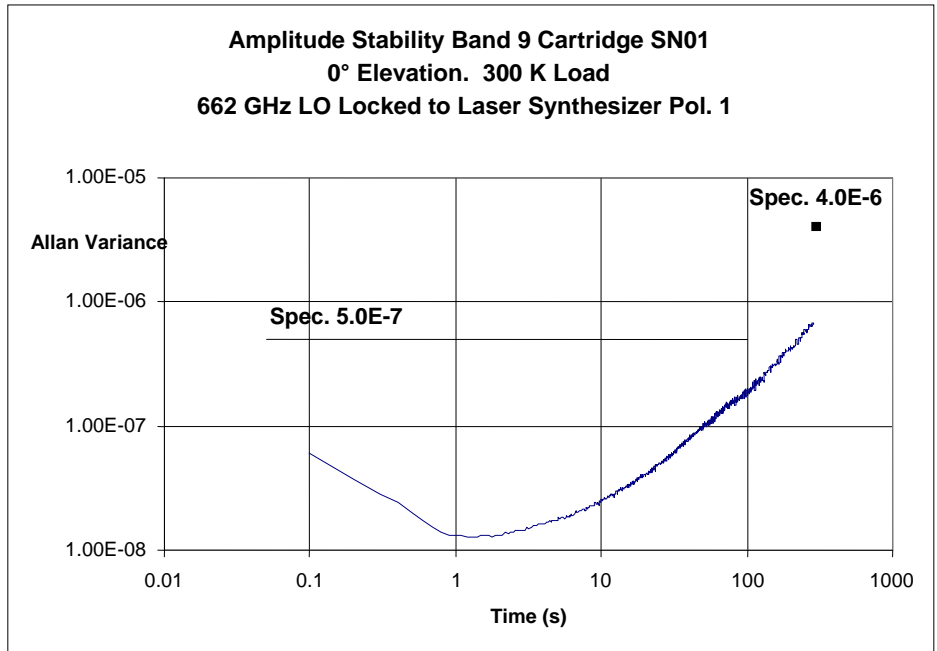

Figure 25 Band 9 Amplitude stability at 0 degrees elevation.

Finally, Far-field beam patterns calculated from near-field scans for the band 9 receiver at 676 GHz are given in Figure 26 (co-polar) and cross polar in Figure 27. The cross polar peak is >22 dB below the co-polar peak at this frequency.

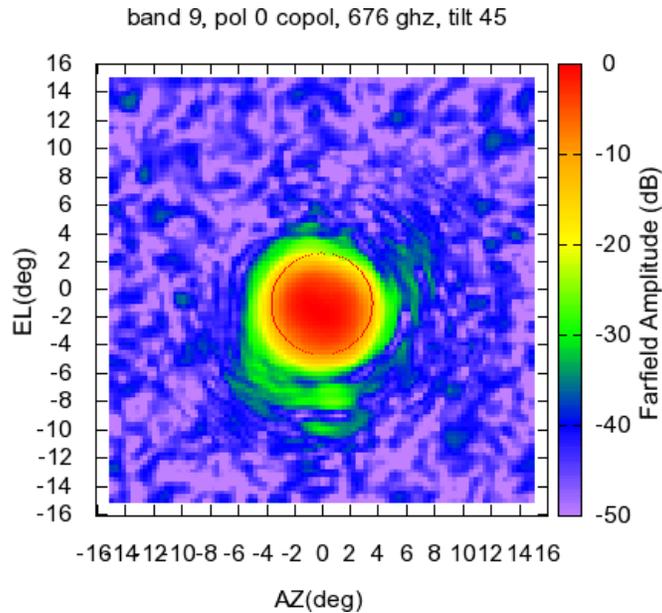

Figure 26 676 GHz Polarization (Pol) 0 co-polar elevation 0. The red circle indicates the size and position of the secondary mirror.

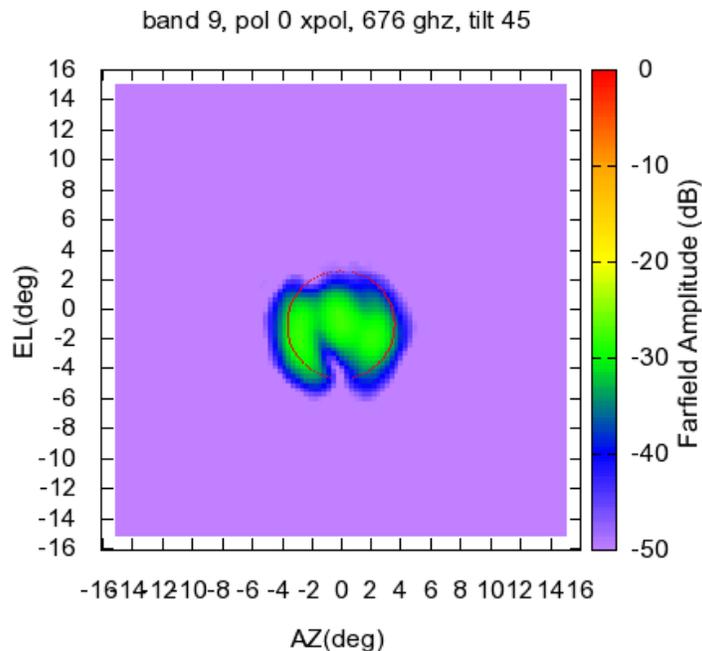

Figure 27 676 GHz Pol 0 cross-pol. elevation 45. The red circle indicates the size and position of the secondary mirror.

From the measured beams the beam efficiencies on the telescope are calculated, as shown in Table 4 and Table 5, as a function of frequency, polarization channel and elevation angle.

Band 9

| Frequency | Polarization | Elevation degrees | Taper efficiency % | spillover efficiency % | Aperture efficiency % |
|---|---|---|---|---|---|
| 620 | 0 | 45 | 91.98 | 94.19 | 86.43 |
| 620 | 1 | 45 | 92.27 | 94.23 | 86.75 |
| 661 | 0 | 0 | 85.63 | 94.79 | 80.93 |
| 661 | 1 | 0 | 85.64 | 94.97 | 81.61 |
| 661 | 0 | 45 | 89.49 | 92.87 | 82.18 |
| 661 | 1 | 45 | 85.42 | 95.58 | 81.29 |
| 700 | 0 | 45 | 84.48 | 96.33 | 80.15 |
| 700 | 1 | 45 | 85.17 | 96.88 | 82.14 |

Table 4 Beam efficiencies.

Taking the difference in x, y positions in the best focal plane for each frequency with a plate scale factor of 2.148 arcsec per mm (for the ALMA 12m antennas), the beam squint between the two polarizations (in beam widths) on the sky is also given in the Table 5. The FWHM for band 9 is 9.36 arcsec at 661 GHz. The X and Y positions can be measured to +/- 100 microns (0.2 arc seconds on the sky) at all frequencies, by rotating the source to 45 degrees to each polarization and measuring the two co-polar scans without further rotation of the source.

| Frequency | Polarization | Elevation degrees | Peak cross polar level | Integrated Polarization efficiency % over subreflector | Beam squint (% of FWHM) |
|---|---|---|---|---|---|
| 620 | 0 | 45 | - | - | - |
| 620 | 1 | 45 | -15.95 | 99.66 | 4.76 |
| 661 | 0 | 0 | -17.38 | 98.95 | - |
| 661 | 1 | 0 | -17.79 | 99.52 | 2.63 |
| 661 | 0 | 45 | -15.98 | 98.81 | - |
| 661 | 1 | 45 | -22.52 | 99.26 | 1.61 |
| 700 | 0 | 45 | -23.94 | 86.16 | - |
| 700 | 1 | 45 | - | - | 7.14 |

Table 5 Cross polar performance and squint.

## V. Conclusion

A complex measurement system has been designed and built to measure a series of important receiver parameters (noise temperature, image rejection, amplitude and phase stability, beam patterns, etc.) for the ALMA Front Ends. The system can make these measurements for all ALMA bands (33 GHz to 960 GHz). System stability and initial receiver measurements have been given.

## VI. Acknowledgements

We wish to thank C. Cunningham for his help with some of the documentation, K. Saini for his help with the design of the phase drift system, and all the members of the various cartridge groups for their help with the operation of their cartridges.

## VII. References

[1] H. Rudolf. "The ALMA Front End." presented at the German Microwave Conf., Karlsruhe, Germany, Mar. 28–30, 2006. http://duepublico.uni-duisburg-essen.de/servlets/DerivateServlet/Derivate-14694/Final_Papers/GM0003-F.pdf.

[2] M.C. Carter, A. Baryshev, M. Harman, B. Lazareff, J. Lamb, S. Navarro, D. John, A. –L. Fontana, G.A. Ediss, C.Y. Tham, S. Withington, F. Tercero, R. Nesti, G.-H. Tan, Y. Sekimoto, M. Matsunaga, H. Ogawa, and S. Claude. "ALMA front-end optics." Ground-based Telescopes. Edited by Oschmann, Jacobus M., Jr. Proceedings of the SPIE, Volume 5489, pp. 1074-1084 (2004).

[3] S. Claude, F. Jiang, P.Niranjanan, P. Dindo, D. Erickson, K. Yeung, D. Derdall, D. Duncan, D. Garcia, D. Henke, B. Leckie, M. Pfleger, G. Rodrigues, K. Szeto, P. Welle, I. Wood, K. Caputa, A. Lichtenberger and S-K. Pan "Performance of the Band 3 (84-116 GHz) receiver for ALMA," Proc. 17th Int. Symp. on Space Terahertz Technology (2006), Paris, pp. 154-157.

[4] G. A. Ediss, J. E. Effland, W. Grammer, N. Horner, A. R. Kerr, D. Koller, E. F. Lauria, S.-K. Pan, M. Sullivan, J. Chen and M. Carter, "ALMA band 6 prototype cartridge: Design and performance," presented at the 15th Int. Symp. on Space Terahertz Technology, Northampton, MA, Apr. 27–27, 2004.

[5] A.M. Baryshev, R. Hesper, F.P. Mena, B.D. Jackson, J. Adema, H. Schaeffer, J. Barkhof and W. Wild, "Design and performance of the 600–720 GHz ALMA band 9 cartridge," 17th International Symposium on Space Terahertz Technology (2006) Paris, pg 89.

[6] P.W. Bond and G.A. Ediss. "Design, alignment and calibration requirements for a sub-millimeter wave frequency tiltable lightweight scanner." Proceedings AMTA 2007, St Louis.

[7] A. R. Kerr, S.-K. Pan, and J. E. Effland. "Sideband Calibration of Millimeter-Wave Receivers." ALMA Memo 357 http://www.alma.nrao.edu/memos/html-memos/alma357/memo357.pdf